\documentstyle[12pt,epsf]{article}
\newcommand{\beq}{\begin{equation}}
\newcommand{\eeq}{\end{equation}}
\newcommand{\bea}{\begin{eqnarray}}
\newcommand{\eea}{\end{eqnarray}}
\newcommand{\beas}{\begin{eqnarray*}}
\newcommand{\eeas}{\end{eqnarray*}}
\newcommand{\nn}{\nonumber}
\newcommand{\limit}{\rightarrow}

\newcommand{\tr}{{\rm tr}}

\newcommand{\bra}{\langle}
\newcommand{\ket}{\rangle}

\begin{document}
\topmargin 0pt
\oddsidemargin 5mm
\headheight 0pt
\topskip 0mm
\begin{titlepage}
\begin{flushright}
May 2001 \\
SACLAY-SPHT/T01/057 \\ 
hep-th/0105284
\end{flushright}
\vspace{1cm}
\begin{Large}
\begin{center}
{\sc Gaussian and Mean Field Approximations}\\
{\sc for Reduced 4D Supersymmetric Yang-Mills Integral} \\
\end{center}
\end{Large}
\vspace{5mm}
\begin{center}
\begin{large}
{\sc Fumihiko Sugino}\footnote{E-mail address: 
{\tt sugino@spht.saclay.cea.fr}\label{Sugino}} \\
\end{large}
\vspace{4mm}
${}^{\ref{Sugino}}${\it Service de Physique Th\'{e}orique, C.E.A. Saclay,}\\
{\it F-91191 Gif-sur-Yvette Cedex, France}\\

\vspace{3cm}
\begin{large} 
Abstract 
\end{large}
\end{center}

   In this paper, we consider a reduced supersymmetric Yang-Mills integral  
with four supercharges by using 
a Gaussian approximation scheme and its improved version. 
We calculate the free energy and the expectation values of 
Polyakov loop and Wilson loop operators by extending the method employed 
in the bosonic case in the previous paper. 
Our results nicely match to the exact and the numerical results obtained before. 
The loop amplitudes exhibit good scaling behaviors similarly as in the bosonic case. 
The 't Hooft like large $N$ limit leads simple formulas 
for the case of the loop length smaller.  
Also, the Polyakov loop and the Wilson loop are computed for the case of the 
loop length sufficiently large, 
where we see that the behavior of the Wilson loop reproduces the result simulated 
for a few smaller values of $N$ at least qualitatively.

\end{titlepage}


\section{Introduction}

In this paper, we consider a supersymmetric Yang-Mills integral, 
which is the complete dimensional reduction of the ${\cal N}=1$ 
supersymmetric Yang-Mills theory on the Euclidean four-dimensions 
with the gauge group $SU(N)$. The classical action is 
$S=S^{(B)}+S^{(F)}$, where\footnote{This fermionic action $S^{(F)}$ 
differs from the action defined in 
eq. (2.1) in ref. \cite{Ambjorn} by the factor $i$. 
Unfortunately, the proof of the positivity of the fermion determinant ${\cal M}$ 
in \cite{Ambjorn} 
has a minor error. 
The identity $\sigma_2{\cal M}\sigma_2={\cal M}^*$ does not hold because 
$A_{a\mu}$ defined in Appendix A in \cite{Ambjorn} is not real. 
However, the outline of the proof is correct. 
In fact, starting with our action and repeating the argument presented there, 
we can prove the positivity. 
Also, in the $N=2$ case we are able to check it 
by an explicit calculation of the fermion determinant, 
which gives the answer identical to eq. (16) in the first paper of ref. \cite{Krauth2}. 
Eventually, since the authors in \cite{Ambjorn} calculate everything 
by using the hermitian matrix 
${\cal D}={\cal M}^{\dagger}{\cal M}$, the factor $i$ does not contribute and 
their final results are correct.}    
\beq
S^{(B)}=-\frac{1}{4g^2}\,\tr\, [X_{\mu}, X_{\nu}]^2, \quad 
S^{(F)}=-\frac{i}{g^2}\,\tr\, 
\bar{\psi}_{\alpha}(\Gamma_{\mu})_{\alpha\beta}[X_{\mu}, \psi_{\beta}]. 
\label{definition}
\eeq
The variables $X_{\mu}$ $(\mu=1,\cdots, 4)$ are $N\times N$ traceless hermitian matrices, 
and $\psi_{\alpha}$, $\bar{\psi}_{\alpha}$ $(\alpha=1,2)$ are Grassmann valued 
traceless $N\times N$ complex matrices. For the Gamma matrices $\Gamma_{\mu}$ 
we take a representation 
$\Gamma_{k}=i\sigma_{k}$, $\Gamma_4={\bf 1}_2$, with $\sigma_k$ ($k=1,2,3$) 
being the Pauli matrices. 
The partition function is defined as 
\beq
Z=\int (dX) (d\psi d\bar{\psi})\, e^{-S}
\eeq
with the integration measures 
\beq
(dX)=\prod_{\mu=1}^4 \prod_{A=1}^{N^2-1}\frac{dX_{\mu}^A}{\sqrt{2\pi}}, \quad 
(d\psi d\bar{\psi})=\prod_{\alpha=1}^2\left(\prod_{A=1}^{N^2-1}
d\psi_{\alpha}^A d\bar{\psi}_{\alpha}^A\right). 
\eeq
The variables are expanded by a basis $T^A$ normalized by $\tr\, 
(T^AT^B)=\frac12\delta^{AB}$. 

 The extended version of this model with sixteen supercharges 
has been proposed as a nonperturbative definition of IIB superstring theory \cite{IKKT}. 
It has no expansion parameters because the coupling constant $g$ 
in the action can be absorbed away into 
an appropriate rescaling of the variables. 
So it is hard to analyze it, as usual, in the form of an expansion of some parameters. 
Under this situation, 
numerical calculations have been performed 
vigorously \cite{Krauth2,Ambjorn,Ambjorn2,Egawa,Bialas}. 
Analytic exact results for the partition function and the correlator of some operators 
have been obtained by using the cohomological field theory 
approach based on 
supersymmetry \cite{MNS,KS,Austing}\footnote{For analytical results from other methods, 
see \cite{Suyama,Krauth,Austing2}. Also, for the cohomological field theory approach 
to quantum mechanical 
and field theoretical matrix models, see \cite{Porrati,Sugino,HKK,Nekrasov}.}.   

 In this paper, we perform analytic calculations expanding the methods of 
Gaussian and mean field approximations, discussed in the bosonic case in ref. \cite{OS}, 
into the 
supersymmetric case. In the bosonic case, the partition function and various correlators 
including Polyakov 
and Wilson loops have been computed. The results for the Gaussian approximation fit better 
to the known results obtained numerically or analytically 
as increasing the number of the variables 
(the rank of the gauge group $N$ or the dimension $D$). In particular, 
it can be seen that the results for the Polyakov and Wilson loops nicely 
reproduce the numerical results 
in ref. \cite{Ambjorn} for the region of the loop length $L$ smaller. 
Also, we tried evaluating the asymptotic behavior of the loop amplitudes 
in the limit $L\limit\infty$ 
by improving the method. Here, we perform a similar analysis 
for the supersymmetric case.   
For various supersymmetric quantum mechanical systems, 
the method of the Gaussian approximation has been 
examined in refs. \cite{KL,KLL}, which suggests that 
it is necessary to formulate the Gaussian 
action so as to respect supersymmetry. 
This treatment is crucial, in particular, 
for observing the dynamical supersymmetry breaking from the behavior 
of the free energy. This is a reason why the Gaussian action is considered 
in the superfield formalism 
including auxiliary fields in refs. \cite{KL,KLL}. 
We follow this lesson and consider the supersymmetric Gaussian action. 
In our case ($D=4$), however 
we can easily realize the supersymmetry without using auxiliary fields. 
The situation is special for the four dimensions, 
so we will have to introduce auxiliary fields 
in the case of $D=6$ or 10 (with eight or sixteen supercharges) \cite{Sugino2}. 

 This paper is organized as follows. 
In the next section, we consider the Gaussian approximation 
to the supersymmetric Yang-Mills 
integral by extending the method used in the bosonic case \cite{OS} 
so as to respect supersymmetry. 
We obtain the results for the partition function and the amplitudes 
of Polyakov and Wilson loops, 
which are reliable for $N$ larger and for the loop length $L$ smaller. 
In section 3, in order to get some insight for behaviors of 
the loop amplitudes for $L$ large, 
we consider an improved version of the Gaussian approximation. 
The last section is devoted to summarize our results 
and mention possible future directions.


\section{Gaussian Approximation}
\setcounter{equation}{0}

  We consider the Gaussian approximation for the supersymmetric 
Yang-Mills integral (\ref{definition}). 
As discussed in refs. \cite{KL,KLL}, respecting supersymmetry 
is crucial to realize a reasonable approximation. 
We start with the following Gaussian action: $S_0=S_0^{(B)}+S_0^{(F)}$,  
\beq
 S_0^{(B)}=\frac{1}{\sigma^2}\tr\,(X_{\mu})^2, \quad 
S_0^{(F)}=\frac{1}{\tau^2}\varepsilon_{\alpha\beta}\,\tr\,(\psi_{\alpha}\psi_{\beta}
-\bar{\psi}_{\alpha}\bar{\psi}_{\beta}). 
\eeq
This is supersymmetric, and the bosonic degrees of freedom 
and the fermionic ones are 
balanced. In fact, it is easy to see that 
the expectation value $\bra S_0\ket_0$ under the classical action $S_0$ vanishes. 
In refs. \cite{KL,KLL}, for the Gaussian approximation 
of supersymmetric quantum mechanical 
systems it is emphasized introducing auxiliary fields in order to balance the 
bosonic and fermionic degrees of freedom in Gaussian actions. 
It is equivalent to matching the freedom in the off-shell level 
because in the Gaussian action 
auxiliary fields contribute to the freedom as well as the dynamical variables. 
This is a reason why the Gaussian action is considered starting 
with the superfield formalism 
in refs. \cite{KL,KLL}. 
It is noted that in our zero-dimensional case 
the bosonic and fermionic degrees of freedom are already 
matched without introducing auxiliary fields.  
In this approximation, the partition function (the free energy $F$) and 
the correlator of the operator ${\cal A}$ are expanded as 
\bea
F & = & F_0-\sum_{n=1}^{\infty}\frac{(-1)^n}{n!}\bra(S-S_0)^n\ket_{C,0},  
\label{cumulantF} \\
\bra {\cal A}\ket & = & \bra{\cal A}\ket_0+
\sum_{n=1}^{\infty}\frac{(-1)^n}{n!}\bra(S-S_0)^n{\cal A}\ket_{C,0},
\label{cumulantO}
\eea
where $F_0$ and $\bra\cdots\ket_0$ denote the free energy and 
the expectation value in the Gaussian 
theory. The suffix $C$ means the connected part of the correlator. 

\subsection{Partition Function}
First we consider the partition function in the Gaussian approximation. 
The outline is same as in the bosonic case \cite{OS}, 
but in the supersymmetric case 
to get a reasonable result 
we have to reorganize the series of $S-S_0$ in (\ref{cumulantF}) 
so that the expansion respects the supersymmetry. 
In order to do so, we introduce a number we will call ``order'' : 
1 assigned  to $S^{(B)}$, $S_0^{(B)}$ and 
$S_0^{(F)}$, and 1/2 to $S^{(F)}$. 
It is easy to see from a diagramatic consideration 
as for the gap equations in ref. \cite{KL} that  
the expansion with respect to the total number of the order is 
natural from the point of view 
of the supersymmetry of the Gaussian action $S_0$. 
Straightforward calculations lead 
\bea
F_0 & = & -2(N^2-1)\ln \frac{\sigma^2}{\tau^2} \nn \\
\bra S^{(B)}-S_0\ket_0 & = & \frac{3\sigma^4}{2g^2}N(N^2-1) \nn \\
-\frac12\bra(S^{(F)})^2\ket_{C,0} & = & -\frac{1}{g^4}N(N^2-1)\sigma^2\tau^4 \nn \\
-\frac12\bra(S^{(B)}-S_0)^2\ket_{C,0} & = & -\frac{45\sigma^8}{16g^4}N^2(N^2-1)+
\frac{3\sigma^4}{g^2}N(N^2-1) \nn \\
\frac12\bra(S^{(B)}-S_0)(S^{(F)})^2\ket_{C,0} & = & \frac{3\sigma^6\tau^4}{2g^6}N^2(N^2-1)
-\frac{3\sigma^2\tau^4}{g^4}N(N^2-1) \nn \\
-\frac{1}{4!}\bra(S^{(F)})^4\ket_{C,0} & = & \frac{5\sigma^4\tau^8}{8g^8}N^2(N^2-1), 
\eea
where the order 1 contributions are the second and third lines, 
and the order 2 contributions the forth and lowers.  
Considering the gap equations from the contribution up to the order 1, i.e. 
\bea
\frac{\partial}{\partial \sigma^2}\left[
F_0+\bra S^{(B)}-S_0\ket_0-\frac12 \bra (S^{(F)})^2\ket_{C,0}\right]=0,   \\ 
\frac{\partial}{\partial \tau^2}\left[
F_0+\bra S^{(B)}-S_0\ket_0-\frac12 \bra (S^{(F)})^2\ket_{C,0}\right]=0,  
\eea
the parameters are determined as 
\beq
\sigma^2=\sqrt{\frac{g^2}{N}}, \quad \tau^2=\left(\frac{g^6}{N}\right)^{1/4}. 
\label{sigma-tau}
\eeq
Then the Gaussian approximation for the free energy up to the order 2 is 
\beq
F=\frac12 (N^2-1)\left[\ln (Ng^2)+1-\frac{11}{8}+
(\mbox{corrections of (order)}\ge 3) \right], 
\label{Gaussian-F}
\eeq
where the second and third terms represent 
the corrections of the order 1 and 2, respectively. 

Let us compare this result with the exact one \cite{Krauth2,MNS}. 
The exact result is given by 
\beq
Z_{\rm exact}=(2g^2)^{-\frac12(N^2-1)}{\cal F}_N\frac{1}{N^2}, 
\quad {\cal F}_N=\frac{2^{N(N+1)/2}\pi^{(N-1)/2}}{2\sqrt{N}\prod_{i=1}^{N-1}i!}, 
\eeq
and the large-$N$ expanded form of the free energy becomes 
\beq
F_{\rm exact}=-\ln Z_{\rm exact}=
\frac12(N^2-1)\left[\ln (Ng^2)-\frac32+\frac{35}{6}
\frac{\ln N}{N^2}+O(N^{-2})\right]. 
\label{expansion-exact}
\eeq
Taking the ratio $\Delta\equiv (F-F_{\rm exact})/F_{\rm exact}$ 
for various values of $N$, 
we have $\Delta(N=3)=-1.10408$, $\Delta(N=4)=-2.19021$, 
$\Delta(N=5)=-3.8156$, $\Delta(N=6)=-8.05504$, 
$\Delta(N=100)=0.464781$, $\Delta(N=1000)=0.23609$. 
Here we put $g^2=1/2$ for the same setting as 
in the bosonic case (eq. (4.6) in ref. \cite{OS}). 
Note that differently from the bosonic case 
the inequality $F\le F_0+\bra S-S_0\ket_0$ no longer holds 
because the integrals of Grassmann variables do not preserve the inequality. 
Thus $\Delta$ can take a negative value. 
Although for the smaller values of $N$ ($N=3$,4,5,6) 
the approximation does not make sense, 
for the larger values ($N=100$, 1000) 
it becomes better as $N$ increases. 
Also, for $N$ larger it can be seen by comparing (\ref{Gaussian-F}) with 
the expansion (\ref{expansion-exact}). 
On the other hand, in the bosonic case (see eq. (4.6) in ref. \cite{OS}) 
the approximation 
gives a reasonable result already at the smaller $N$. 
We can say that in the supersymmetric case 
the convergence of the Gaussian approximation for the partition function 
with respect to $N$ is slower than that in the bosonic case.     

\subsection{Polyakov Loop}

  Next we consider the expectation value of the Polyakov loop operator 
$\hat{P}(L)=\frac1N\tr\,(e^{iLX_1})$. 
In the expansion (\ref{cumulantO}), 
calculation of the correlators $\bra\hat{P}(L)\ket_0$, 
$-\bra\hat{P}(L)(S-S_0)\ket_{C,0}$ and 
$\frac12\bra\hat{P}(L)(S^{(B)}-S_0)^2\ket_{C,0}$ reduces 
to that of the purely bosonic contributions, 
and leads to the same formulas as obtained in the bosonic case \cite{OS}, 
due to decoupling of the fermionic contributions or to the parity symmetry 
$X_{\mu}\limit -X_{\mu}$ of the Gaussian action. 
So our task up to the order 2 reduces essentially 
to evaluating the three correlators 
$\frac12\bra\hat{P}(L)(S^{(F)})^2\ket_{C,0}$, 
$-\frac12\bra\hat{P}(L)(S^{(B)}-S_0)(S^{(F)})^2\ket_{C,0}$ and 
$\frac{1}{4!}\bra\hat{P}(L)(S^{(F)})^4\ket_{C,0}$. 
In calculating the fermionic integrals, the formula 
\beq
\bra{\cal B}(X)(S^{(F)})^2\ket_{C,0}=
\frac{\sigma^2\tau^4}{g^4}N\bra{\cal B}(X)S_0^{(B)}\ket_{C,0}
\eeq
is useful. Here ${\cal B}(X)$ is a bosonic operator 
containing no $\psi_{\alpha}$ and $\bar{\psi}_{\alpha}$. 
The results are summarized as 
\begin{itemize}
\item 
The lowest order 
\beq
\bra \hat{P}(L)\ket_0=
e^{-\frac{L^2\sigma^2}{4N}(N-1)}F\left(1-N,2;\frac{L^2\sigma^2}{2}\right), 
\eeq
\item 
The order 1 correction 
\bea
\lefteqn{-\bra\hat{P}(L)(S^{(B)}-S_0)\ket_{C,0}+
\frac12\bra\hat{P}(L)(S^{(F)})^2\ket_{C,0}} \nn \\
& = & \left(\frac{3\sigma^4}{2g^2}N-\frac{\sigma^2\tau^4}{2g^4}N-1\right)
\frac{L^2\sigma^2}{4}
\left[e^{-\frac{L^2\sigma^2}{4N}(N-1)}
\left\{(N+1)F\left(1-N,3;\frac{L^2\sigma^2}{2}\right) \right.\right. 
\nn \\
 & & \left.\left.-F\left(1-N,2;\frac{L^2\sigma^2}{2}\right)\right\}
-\frac1N\bra\hat{P}(L)\ket_0\right], 
\eea
\item
The order 2 correction 
\bea
\lefteqn{\frac12\bra\hat{P}(L)(S^{(B)}-S_0)^2\ket_{C,0}
-\frac12\bra\hat{P}(L)(S^{(B)}-S_0)(S^{(F)})^2\ket_{C,0}
+\frac{1}{4!}\bra\hat{P}(L)(S^{(F)})^4\ket_{C,0}} \nn \\
 & = & e^{-\frac{L^2\sigma^2}{4N}(N-1)}\left[
-\left(\frac{3\sigma^4}{2g^2}N-\frac{\sigma^2\tau^4}{2g^4}N-1\right)^2+
\frac{N-3}{4}\left(\frac{3\sigma^8}{g^4}
-\frac{\sigma^4\tau^8}{2g^8}\right)\right]\frac{L^4\sigma^4}{32N}
\nn \\
& & \times\left[\left(\begin{array}{c}N \\ 4 \end{array}\right) 
F\left(4-N,5;\frac{L^2\sigma^2}{2}\right)
+\left(\begin{array}{c}N+1 \\ 4 \end{array}\right) 
F\left(3-N,5;\frac{L^2\sigma^2}{2}\right)\right] 
\nn \\
 & + & e^{-\frac{L^2\sigma^2}{4N}(N-1)}\left[
\left(\frac{3\sigma^4}{2g^2}N-
\frac{\sigma^2\tau^4}{2g^4}N-1\right)^2+
\frac{N+3}{4}\left(\frac{3\sigma^8}{g^4}-
\frac{\sigma^4\tau^8}{2g^8}\right)\right]\frac{L^4\sigma^4}{32N}
\nn \\
& & \times\left[\left(\begin{array}{c}N+2 \\ 4 \end{array}\right) 
F\left(2-N,5;\frac{L^2\sigma^2}{2}\right)
+\left(\begin{array}{c}N+3 \\ 4 \end{array}\right) 
F\left(1-N,5;\frac{L^2\sigma^2}{2}\right)\right] 
\nn \\ 
& + & e^{-\frac{L^2\sigma^2}{4N}(N-1)}\left(-\frac{3\sigma^8}{g^4}+
\frac{\sigma^4\tau^8}{2g^8}\right)
\frac{L^4\sigma^4}{32N}\left[\left(\begin{array}{c} N \\ 3 \end{array}\right)
F\left(3-N,4;\frac{L^2\sigma^2}{2}\right)\right. \nn \\
 &  & \left.+\left(\begin{array}{c} N+1 \\ 3 \end{array}\right)
F\left(2-N,4;\frac{L^2\sigma^2}{2}\right) 
+\left(\begin{array}{c} N+2 \\ 3 \end{array}\right)
F\left(1-N,4;\frac{L^2\sigma^2}{2}\right)\right] \nn \\
 & + & e^{-\frac{L^2\sigma^2}{4N}(N-1)}
\left(\begin{array}{c}N \\ 2 \end{array}\right)
\left[\left\{-\left(\frac{9\sigma^4}{4g^2}N-1\right)^2
-\frac{3\sigma^8}{16g^4}N(3N-2)+\frac{\sigma^2\tau^4}{g^4}N
\left(\frac{9\sigma^4}{4g^2}N-2\right)
\right.\right.\nn \\
 & & \left.\left.+\frac{\sigma^4\tau^8}{16g^8}N(10N-1)\right\}
\frac{L^2\sigma^2}{4N}-\left(\frac{3\sigma^4}{2g^2}N-
\frac{\sigma^2\tau^4}{2g^4}N-1\right)^2
\frac{L^4\sigma^4}{16N^2}\right]
F\left(2-N,3;\frac{L^2\sigma^2}{2}\right) \nn \\
 &+ & e^{-\frac{L^2\sigma^2}{4N}(N-1)}
\left(\begin{array}{c}N+1 \\ 2 \end{array}\right)
\left[\left\{-\left(\frac{9\sigma^4}{4g^2}N-1\right)^2
-\frac{3\sigma^8}{16g^4}N(3N+2)+
\frac{\sigma^2\tau^4}{g^4}N\left(\frac{9\sigma^4}{4g^2}N-2\right)
\right.\right.\nn \\
 & & \left.\left.+\frac{\sigma^4\tau^8}{16g^8}N(10N+1)\right\} 
\frac{L^2\sigma^2}{4N}-
\left(\frac{3\sigma^4}{2g^2}N-\frac{\sigma^2\tau^4}{2g^4}N-1\right)^2
\frac{L^4\sigma^4}{16N^2}\right]
F\left(1-N,3;\frac{L^2\sigma^2}{2}\right) \nn \\
 &+ & \left[\left\{\left(\frac{9\sigma^4}{4g^2}N-1\right)^2+
\frac{15\sigma^8}{16g^4}N^2
-\frac{\sigma^2\tau^4}{g^4}N\left(\frac{9\sigma^4}{4g^2}N-2\right)-
\frac{11\sigma^4\tau^8}{16g^8}N^2\right\}
\frac{L^2\sigma^2}{4N} \right. \nn \\
 & & \left. +\left(\frac{3\sigma^4}{2g^2}N-
\frac{\sigma^2\tau^4}{2g^4}N-1\right)^2\frac{L^4\sigma^4}{32N^2}
\right]\bra\hat{P}(L)\ket_0,  
\eea
\end{itemize}
where $F$ is the confluent hypergeometric function: 
$F(\alpha,\beta;z)=
\sum_{n=0}^{\infty}
\frac{\alpha(\alpha+1)\cdots(\alpha+n-1)}{\beta(\beta+1)\cdots(\beta+n-1)}
\frac{z^n}{n!}.$   
Let us consider the value at the solution (\ref{sigma-tau}). 
Then, the order 1 correction vanishes, which is the same situation as in 
the bosonic case. Also, the 't Hooft like limit $N\limit \infty$ 
with $Ng^2\equiv G^2$ fixed leads to 
the simple formula expressed by the Bessel functions: 
\beq
\bra\hat{P}(L)\ket=\frac{1}{\sqrt{\alpha}}J_1(2\sqrt{\alpha})
+\frac{5}{16}\left[J_4(2\sqrt{\alpha})-4J_2(2\sqrt{\alpha})\right]+
(\mbox{corrections of (order)}\ge 3) 
\eeq
where $\alpha=L^2G/2$. In ref. \cite{Ambjorn} 
some numerical results about Polyakov and 
Wilson loops are reported. 
We can compare our result with the numerical one presented there. 
Let us take $Ng^2=48$ and consider the quantity for various values of 
$N$ keeping $Ng^2$ fixed, 
which gives the same setting as in ref. \cite{Ambjorn}. 
Figure 5 in ref. \cite{Ambjorn} is the result to be compared with 
ours\footnote{There is a typo in 
the horizontal axis in Figure 5 in ref. \cite{Ambjorn}. 
The horizontal axis represents $k/\sqrt{g}$ 
not $k^2/g$. I thank authors of ref. \cite{Ambjorn} for informing me of this fact.}. 
The variable $k/\sqrt{g}$ in the horizontal axis there corresponds to $L$ in our setting. 
We plot our results up to the order 2 for various values of $N$ in Fig. 1. 
\begin{figure}
\epsfxsize=9cm \epsfysize=6cm
\centerline{\epsfbox{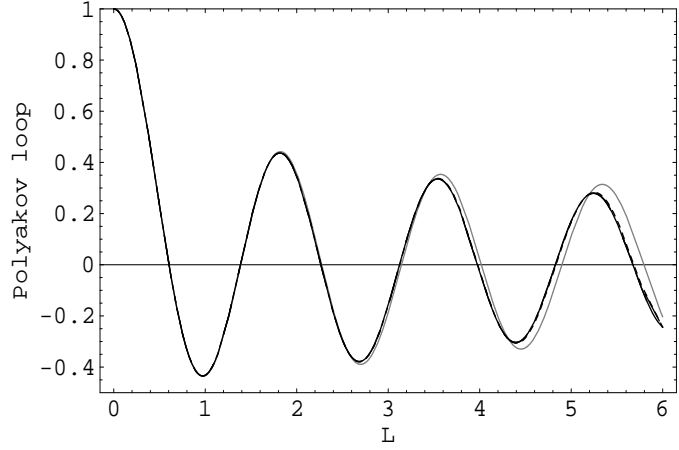}}
\caption{Polyakov loops up to the order 2 
in the case $N=16$, 48, 100, 400 and $\infty$.
The gray line shows the case $N=16$, the dotted line $N=48$, 
the dashed line $N=100$, the dot-dashed line $N=400$, and the solid line 
$N=\infty$.}
\end{figure}
The result exhibits a quite nice scaling behavior in the region $0<L<6$ 
in particular for $N\ge 48$ so that the curves lie almost on top of each other. 
In Fig. 2 we show that the approximation gets better 
as increasing the order from 1 to 2. 
Comparing the curves for the order 1 (the dotted line) 
and the order 2 (the solid line) 
with the numerical result (the gray dots), 
we see that the order 2 result nicely reproduces 
the numerical result for the region of $L$ smaller (up to about 0.6)\footnote{The exact result 
of the Polyakov loop amplitude is non-analytic at $L=0$ 
because its second derivative $\bra\frac1N\tr (X_1)^2\ket$ is logarithmically 
divergent \cite{Austing2}. The Gaussian approximation up to the order 2 
can not reproduce this non-analyticity, but the approximation is considered to get 
better in the 't Hooft limit from the observation that 
the non-analyticity vanishes in the 't Hooft limit \cite{Ambjorn3}.}  
For the region of $L$ larger 
the contributions of the higher orders are considered to become more important. 
For the order 2 result in the region of $L$ larger, 
the amplitude of the oscillation is 
larger compared to the bosonic case (the gray line). 
Agreement for the region seems to be not so better than 
that in the bosonic case 
as long as comparing the results up to the order 2. 
\begin{figure}
\epsfxsize=9cm \epsfysize=6cm
\centerline{\epsfbox{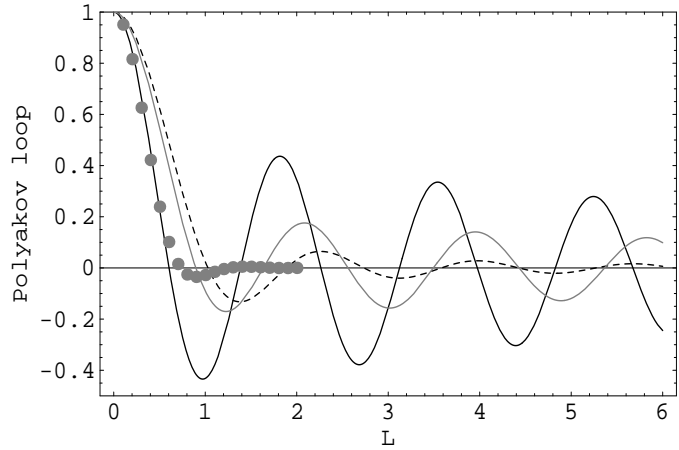}}
\caption{Polyakov loop in the case $N=\infty$ up to the order 1 (the 
dotted line) and up to the order 2 (the solid line). 
The gray dots indicate some points of the numerical data 
for the $N=48$ case in Figure 5 in ref. \cite{Ambjorn}.  
The gray line is the result of the Gaussian approximation 
(up to the second order) for the bosonic case $(D,N)=(4,\infty)$ 
in Figure 3 in ref. \cite{OS}. }
\end{figure}

\subsection{Wilson Loop}

We calculate the expectation value of the Wilson loop 
$\hat{W}(L)=\frac1N\tr\,(e^{iLX_1}e^{iLX_2}e^{-iLX_1}e^{-iLX_2})$ 
by the Gaussian approximation up to the order 1. 
The formulas for $\bra \hat{W}(L)\ket_0$ and 
$-\bra\hat{W}(L)(S-S_0)\ket_{C,0}$ 
are same as the ones obtained 
in the bosonic case \cite{OS} (eqs. from (4.28) to (4.37) in the paper). 
Thus the quantity we have to evaluate is only 
$\frac12\bra\hat{W}(L)(S^{(F)})^2\ket_{C,0}$, 
the result for which is 
\beq
\frac12\bra\hat{W}(L)(S^{(F)})^2\ket_{C,0}=
-\frac{\tau^4}{g^4}\frac{N^4}{N^2-1}\tilde{f}_2(L)
(-1+u(L)). 
\eeq
We used the functions $\tilde{f}_2(L)$ and $u(L)$ defined in ref. \cite{OS} 
(See eqs. (B.10) and (B.11) in the paper). 
The explicit formula is too lengthy to write down directly. 
However, in the 't Hooft like limit it becomes remarkably simplified as 
\bea
\lefteqn{\bra\hat{W}(L)\ket = 
1-\left(\frac{1}{\alpha}(J_1(2\sqrt{\alpha}))^2-1\right)^2}\nn \\
 & & +\frac{1}{\alpha^2}\left(9(J_1(2\sqrt{\alpha}))^2(J_3(2\sqrt{\alpha}))^2+
8(J_2(2\sqrt{\alpha}))^4
\right)+(\mbox{corrections of (order)}\ge 2). 
\eea 
In Fig. 3, we plot the behavior for various values of $N$ 
with $Ng^2=48$ being fixed. 
A nice scaling behavior for $N\ge 48$ can be seen 
as well as in the bosonic case \cite{OS}. 
\begin{figure}
\epsfxsize=9cm \epsfysize=6cm
\centerline{\epsfbox{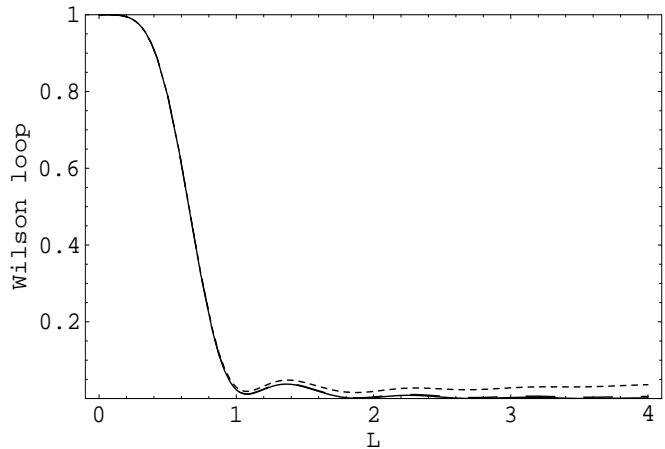}}
\caption{Wilson loops up to the order 1 
for $N=16$, 48 and $\infty$. 
The dotted line shows the case $N=16$, the dashed line $N=48$, 
and the solid line $N=\infty$. 
The scaling is so good that the curves for the $N=48$ and $\infty$ cases are on top of 
each other.}
\end{figure}
Our result can be compared with the numerical result 
in Figure 3 in ref. \cite{Ambjorn}.  
Fig. 4 shows our result with some data for $N=48$ case 
in Figure 3 in ref. \cite{Ambjorn}. 
Taking into account the order 1 correction, 
the fitting of our result to the numerical one 
becomes better for the region of $L$ smaller (up to about 0.7). 
Also, for the Wilson loop the behavior (the solid line) does not exhibit 
major discrepancies from 
that in the bosonic case (the gray line).  
\begin{figure}
\epsfxsize=9cm \epsfysize=6cm
\centerline{\epsfbox{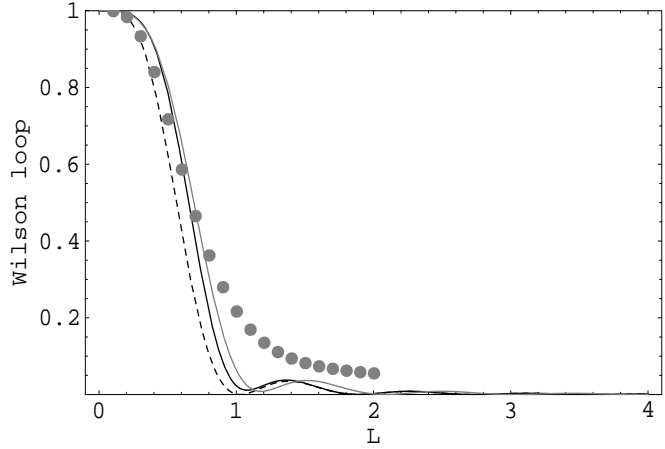}}
\caption{Wilson loop in the case $N=\infty$ up to the leading order 
(the dotted line) and up to the order 1 (the solid line). 
The gray dots indicate some points of the numerical result for $N=48$ 
in Figure 3 in ref. \cite{Ambjorn}. 
The gray line, the result of the Gaussian approximation 
(up to the first order) for the bosonic case $(D,N)=(4,\infty)$ 
in Figure 6 in ref. \cite{OS}, 
is added for comparison.} 
\end{figure}


\section{Improved Mean Field Approximation}
\setcounter{equation}{0}
   In order to obtain some insight for the behavior of the Polyakov and Wilson loops 
in the region of $L$ large, we consider about the improved mean field approximation 
along the same line as discussed in the bosonic case \cite{OS}.  
First, we begin with obtaining an interpretation 
for the Gaussian approximation as a mean field 
approximation. 
We take the following mean field action: $S_M=S_M^{(B)}+S_M^{(F)}+\frac{1}{\lambda}$, 
where 
\bea
 & & S_M^{(B)}=\frac{1}{\sigma^2}\tr\,(X_{\mu})^2, \quad 
S_M^{(F)}=\frac{1}{\tau^2}\varepsilon_{\alpha\beta}\tr\, (\psi_{\alpha}\psi_{\beta})
-\frac{1}{\bar{\tau}^2}\varepsilon_{\alpha\beta}\tr\, 
(\bar{\psi}_{\alpha}\bar{\psi}_{\beta}), \nn \\
 & & \frac{1}{\sigma^2}=\frac{3}{4g^2}
\frac{N^2}{N^2-1}\bra\frac1N\tr(X_{\mu})^2\ket \nn \\
 & & \quad +\frac{1}{2g^4}\frac{N}{(N^2-1)(N^2-2)}
\varepsilon_{\alpha\beta}\varepsilon_{\gamma\delta}
\left[\bra\tr(\psi_{\alpha}\psi_{\beta})
\tr(\bar{\psi}_{\gamma}\bar{\psi}_{\delta})\ket 
+\bra\tr(\psi_{\alpha}\bar{\psi}_{\gamma})
\tr(\psi_{\beta}\bar{\psi}_{\delta})\ket\right], \nn\\
 & & \frac{1}{\tau^2}=
\frac{1}{2g^4}\frac{1}{(N^2-1)(N^2-2)}\varepsilon_{\alpha\beta}
\left[\bra\tr (X_{\mu})^2\tr(\bar{\psi}_{\alpha}\bar{\psi}_{\beta})\ket 
-\bra\tr(X_{\mu}\bar{\psi}_{\alpha})\tr(X_{\mu}\bar{\psi}_{\beta})\ket\right], \nn\\
 & &   \frac{1}{\bar{\tau}^2}=
-\frac{1}{2g^4}\frac{1}{(N^2-1)(N^2-2)}\varepsilon_{\alpha\beta}
\left[\bra\tr (X_{\mu})^2\tr(\psi_{\alpha}\psi_{\beta})\ket 
-\bra\tr(X_{\mu}\psi_{\alpha})\tr(X_{\mu}\psi_{\beta})\ket\right], 
\label{self-consistency1}
\eea
$\lambda$ is a constant determined later. 
The bosonic mean field action $S_M^{(B)}$ comes from 
the contraction of two bosons in $S^{(B)}$ and 
of four fermions in $\frac12(S^{(F)})^2$. 
The fermionic action $S_M^{(F)}$ from the contraction of 
two bosons and two fermions in $\frac12(S^{(F)})^2$. From the $O(4)$-rotational symmetry 
and the $SU(N)$-gauge symmetry, 
the following index structures are assumed: 
\bea
\bra(X_{\mu})_{ij}(X_{\nu})_{kl}\ket & = & c\delta_{\mu\nu}{\cal I}_{ijkl}, 
\label{index1} \\
\bra(\psi_{\alpha})_{ij}(\psi_{\beta})_{kl}
(\bar{\psi}_{\gamma})_{mn}(\bar{\psi}_{\delta})_{pq}\ket
 & = & \varepsilon_{\alpha\beta}\varepsilon_{\gamma\delta}
\left[a{\cal I}_{ijkl}{\cal I}_{mnpq}
+b{\cal I}_{ijmn}{\cal I}_{klpq}+b{\cal I}_{ijpq}{\cal I}_{klmn}\right], 
\label{index2} \\
\bra(X_{\mu})_{ij}(X_{\nu})_{kl}(\psi_{\alpha})_{mn}(\psi_{\beta})_{pq}\ket
 & = & \delta_{\mu\nu}\varepsilon_{\alpha\beta}
\left[a'{\cal I}_{ijkl}{\cal I}_{mnpq}
+b'{\cal I}_{ijmn}{\cal I}_{klpq}+b'{\cal I}_{ijpq}{\cal I}_{klmn}\right], 
\label{index3}
\eea
where $c$, $a$, $b$, $a'$ and $b'$ are scalar constants with no indices, and 
${\cal I}_{ijkl}\equiv \delta_{il}\delta_{jk}-\frac1N\delta_{ij}\delta_{kl}$. 
We evaluate $\bra\cdots\ket$ in eqs. (\ref{self-consistency1}) 
by replacing it with the expectation value 
under the mean field action $\bra\cdots\ket_M$, 
leaving the self-consistency conditions 
\beq
\frac{1}{\sigma^2} =  \frac{3N}{2g^2}\sigma^2-\frac{N}{2g^4}\tau^4, \quad 
\frac{1}{\tau^2} = \frac{N}{g^4}\sigma^2\tau^2. 
\label{self-consistency2}
\eeq
Here we take $\tau$ to be real 
because we can always absorb the phases of $\tau$ and $\bar{\tau}$ 
by rotating the fermions in $S_M^{(F)}$ as $\psi \limit e^{i\theta}\psi$, 
$\bar{\psi} \limit e^{-i\theta}\bar{\psi}$. 
Eqs. (\ref{self-consistency2}) lead the solution (\ref{sigma-tau}). 
As same in the bosonic case \cite{OS}, $\lambda$ is determined 
by the requirement $Z=Z_M$ 
($Z_M$ is the partition function for the mean field action $S_M$.) as 
\beq
\frac{1}{\lambda}=-\sum_{n=1}^{\infty}\frac{(-1)^n}{n!}\bra(S-S_M')^n\ket_{C,M}, 
\eeq
with $S_M'=S_M^{(B)}+S_M^{(F)}$. Then, the partition function becomes 
\beq
F=-\ln Z=-\ln Z_M=-2(N^2-1)\ln \frac{\sigma^2}{\tau^2}-\sum_{n=1}^{\infty}
\frac{(-1)^n}{n!}\bra(S-S_M')^n\ket_{C,M},
\eeq
that reproduces the Gaussian approximation result in section 2.1. 

 Next, we consider the unnormalized correlator 
$\bra {\cal O}\ket'\equiv\int (dX)(d\psi d\bar{\psi})\, {\cal O}e^{-S}$ 
where ${\cal O}$ is isotropic in the four dimensions and $SU(N)$-invariant. 
We repeat the above argument regarding 
$\bra {\cal O}\ket'$ as the partition function of a theory with the Boltzmann weight 
${\cal O}e^{-S}$. 
Then, for such ${\cal O}$, the correlators 
$$
\bra(X_{\mu})_{ij}(X_{\nu})_{kl}{\cal O}\ket,\quad  
\bra(\psi_{\alpha})_{ij}(\psi_{\beta})_{kl}(\bar{\psi}_{\gamma})_{mn}
(\bar{\psi}_{\delta})_{pq}{\cal O}\ket, \quad  
\bra(X_{\mu})_{ij}(X_{\nu})_{kl}(\psi_{\alpha})_{mn}(\psi_{\beta})_{pq}{\cal O}\ket
$$ 
have the index structures same as the amplitudes without ${\cal O}$ 
in (\ref{index1}), (\ref{index2}) 
and (\ref{index3}),
respectively. 
When starting with the mean field action: $S_M=S_M^{(B)}+S_M^{(F)}+\frac{1}{\tilde{\lambda}}$,  
\beq
S_M^{(B)}=\frac{1}{\tilde{\sigma}^2}\tr (X_{\mu})^2, \quad 
S_M^{(F)}=\frac{1}{\tilde{\tau}^2}\varepsilon_{\alpha\beta}\tr \psi_{\alpha}\psi_{\beta}
-\frac{1}{\bar{\tilde{\tau}}^2}\varepsilon_{\alpha\beta}
\tr \bar{\psi}_{\alpha}\bar{\psi}_{\beta}, 
\eeq
the self-consistency conditions give the formulas for $\tilde{\sigma}$, $\tilde{\tau}$ 
and $\bar{\tilde{\tau}}$ 
which are in the form same as eqs. (\ref{self-consistency1}) with the every correlators 
$\bra\cdots\ket$ replaced to 
$\bra\cdots\ket^{\cal O}_M\equiv\bra\cdots{\cal O}\ket_M/\bra {\cal O}\ket_M$. 
In particular, in the case that ${\cal O}$ depends only on $X_{\mu}$'s, 
which is a situation 
we will consider from now, these conditions read 
\bea
\frac{1}{\tilde{\sigma}^2} & = & \frac{3}{4g^2}\frac{N^2}{N^2-1}
\bra\frac1N\tr(X_{\mu})^2\ket_M^{\cal O}-
\frac{N^2-1}{N}\frac{1}{\bra\frac1N\tr(X_{\mu})^2\ket_M^{\cal O}}, \nn \\
\frac{1}{\tilde{\tau}^4} & = & 
\frac{1}{2g^4}\frac{N^2}{N^2-1}\bra\frac1N\tr(X_{\mu})^2\ket_M^{\cal O}. 
\label{self-consistency3}
\eea
Note that the computation of the quantity $\bra\frac1N\tr(X_{\mu})^2\ket_M^{\cal O}$ 
completely reduces to the one 
in the bosonic case, 
because the fermionic contribution of the mean field action is factorized and cancelled 
in each expectation value $\bra \cdots\ket_M$. 
From the requirement $\bra{\cal O}\ket'=\bra{\cal O}\ket_M'$, 
after the same calculation as in the bosonic case \cite{OS}, 
$\tilde{\lambda}$ is determined as 
\beq
\frac{1}{\tilde{\lambda}}= 
-\sum_{m=1}^{\infty}\frac{(-1)^m}{m!}\bra(S-S_M')^m\ket_{C,M} 
-\ln\left[1+\frac{1}{\bra{\cal O}\ket_M}\sum_{n=1}^{\infty}\frac{(-1)^n}{n!}
\bra{\cal O}(S-S_M')^n\ket_{C,M}\right]. 
\eeq
The expression of the correlator becomes 
\beq
\bra {\cal O}\ket'=
\left(\frac{\tilde{\sigma}^2}{\tilde{\tau}^2}\right)^{2(N^2-1)}e^{-1/\tilde{\lambda}}
\bra {\cal O}\ket_M. 
\label{correlator}
\eeq
Finally, dividing this by 
the partition function evaluated up to the order same as done for 
$\bra {\cal O}\ket'$, 
the formula of the correlator $\bra {\cal O}\ket$ is obtained. 
In the following two subsections, 
we try to evaluate the amplitudes of the Polyakov and Wilson loops 
for the case of $L$ large. 

\subsection{Polyakov Loop}
  
 For the Polyakov loop amplitude, we take the operator ${\cal O}$ as 
${\cal O}=\frac{1}{2\cdot 4}\frac1N \sum_{\mu=1}^4\tr (e^{iLX_{\mu}}+e^{-iLX_{\mu}})$. 
Obviously, if the rotational symmetry of $O(4)$ is not 
sponateously broken\footnote{At least, 
in the case of $N$ finite, this always holds. 
When it is spontaneously broken, if we regard the correlator $\bra\cdots\ket$ 
as the sum of the expectation value at 
each superselection sector, 
the equality $\bra{\cal O}\ket= \bra\hat{P}(L)\ket$ holds. 
The spontaneous breaking of the rotational (Lorentz) symmetry 
in reduced Yang-Mills integrals is 
discussed in refs. \cite{Ambjorn2,Aoki,Hotta,Burda,Ambjorn3,Nishi}.},  
$\bra{\cal O}\ket$ is equal to $\bra\hat{P}(L)\ket$. 
When $x\equiv L^2\tilde{\sigma}^2$  and $y\equiv L^3\tilde{\tau}^2$ large, 
the consistency conditions (\ref{self-consistency3}) 
can be solved in the expanded form: 
\bea
x_{\pm} & = & 
e^{\pm i\pi /3}\left(\frac{16}{3}(N+1)L^4g^2\right)^{1/3}+\frac43N(N+2)+O((L^4g^2)^{-1/3}), 
\nn \\
y_{\pm} & = & 
e^{\pm i\pi /6}(18(N+1))^{1/6}(L^4g^2)^{2/3}\left[1+\frac43N(4N+5)e^{\mp i\pi /3}
\left(\frac{3}{16}\frac{1}{N+1}\frac{1}{L^4g^2}\right)^{1/3}\right. \nn \\
 & & \left.\frac{}{}+O((L^4g^2)^{-2/3}\right]^{1/2}. 
\eea
We do not consider the solution such that ${\rm Re}\, x<0$, because 
it leads to the unphysical result blowing up as $L\limit \infty$. 
Corresponding to the above $x$, $\tilde{\lambda}$ is given by 
\beq
\frac{1}{\tilde{\lambda}}= \left\{\begin{array}{ll}
(N-1)\left(\frac{1}{4N}x+N+O(x^{-1})\right) & \mbox{(up to the order 1)} \\
(N-1)\left(\frac{1}{2N}x+3N+2-\frac{5}{12}(N+1)^2+O(x^{-1})\right) &  
\mbox{(up to the order 2)}. 
\end{array}\right. 
\eeq
If we regard the solution as the sum of the $\pm$ branches 
similarly to the bosonic case \cite{OS}, 
dividing the result corresponding to (\ref{correlator}) 
by the partition function up to the order 1 or 2, we get the final answer 
\bea
\bra\hat{P}(L)\ket & \sim & C_{(n)}(L^4g^2)^{-(N-1)^2/6}\exp\left[-u_{(n)}
\frac{N-1}{N}\left(\frac{16}{3}(N+1)L^4g^2\right)^{1/3}\right] \nn \\
& &\times \cos\left[\sqrt{3}u_{(n)}\frac{N-1}{N}
\left(\frac{16}{3}(N+1)L^4g^2\right)^{1/3}-\frac{\pi}{3}(N-1)(N+2)
\right], 
\eea
where up to the order 1 ($n=1$): $u_{(1)}=1/4$, and up to the order 2 ($n=2$): $u_{(2)}=3/8$. 
The overall constants are 
\beas
C_{(1)} & = & 
\frac{(-1)^{N-1}N^{(N^2-1)/2}}{2^{N-2}N!}(N+1)^{(N-1)(N+2)/3}
\left(\frac{128}{81}\right)^{(N^2-1)/3}
\left(\frac{16}{3}\right)^{(N-1)/3}e^{-(N-1)(7N+5)/6}, \\ 
C_{(2)} & = & 
\frac{(-1)^{N-1}N^{(N^2-1)/2}}{2^{N-2}N!}(N+1)^{(N-1)(N+2)/3}
\left(\frac{128}{81}\right)^{(N^2-1)/3}
\left(\frac{16}{3}\right)^{(N-1)/3}e^{(N^2-1)[\frac{5}{12}(N+1)-\frac{67}{16}]}. 
\eeas
It is interesting to compare this with the exact answer obtained 
in ref. \cite{Krauth}. 
In \cite{Krauth} the eigenvalue distribution 
$\rho(\lambda)=\bra\frac1N\tr\delta(\lambda-X_1)\ket$ is exactly 
evaluated for the $N=2$ case (see eq. (12) in ref. \cite{Krauth}). 
The Polyakov loop is related to this quantity via the Fourier transformation: 
$\bra\hat{P}(L)\ket=\int_{-\infty}^{\infty}d\lambda\, e^{iL\lambda}\rho(\lambda)$. 
This integral can be easily estimated in the case $L$ large, and we end up with 
\beq
\bra\hat{P}(L)\ket = 
-\frac{\sqrt{3\pi}}{\Gamma(\frac52)}\left(\frac{L^4g^2}{2^7}\right)^{1/3}
\exp\left[-\frac32\left(\frac{L^4g^2}{2^7}\right)^{1/3}\right]
\times[1+O((L^4g^2)^{-1/3})]. 
\eeq
Although our result reproduces the power $(L^4g^2)^{1/3}$ of the exponential decay, 
the oscillating behavior can not be seen in the exact result, 
also the power behavior is different between 
the improved mean field result ($(L^4g^2)^{-1/6}$) 
and the exact result ($(L^4g^2)^{1/3}$). 
Moreover, the exponential decay behavior of the improved mean field result goes as 
$\exp\left[-2^{2/3}(2^{-7}L^4g^2)^{1/3}\right]$ for up to the order 1 and 
$\exp\left[-3\cdot 2^{-1/3}(2^{-7}L^4g^2)^{1/3}\right]$ for up to the order 2. 
Since the coefficients are $2^{2/3}=1.58\cdots$ and $3\cdot 2^{-1/3}=2.38\cdots$, 
we can not say that the approximation tends to converge 
to the exact results as increasing the order. 
About this disagreement, we could consider as follows. 
The Gaussian or mean field approximation becomes better 
when the number of variables is larger, which means that $N$ is larger. 
As is seen in the analysis of the partition function in section 2.1, 
the result begins exhibiting the nice convergence 
for $N$ larger than about 100. For smaller $N$ ($N=O(1)$), 
the approximation does not make sense. 
There will be a possibility 
that the Polyakov loop has the same tendency as the partition function 
for the dependence of $N$ 
and that our result makes sense for $N$ large.  
In the result of the Gaussian approximation in section 2.2, 
because all the confluent hypergeometric functions appearing there 
are polynomials of $L^2\sigma^2$ for $N\ge 4$, 
the oscillating behavior terminates at some finite value of $L$.  
When $N$ increases, the degree of the polynomials increases and 
the oscillating behavior continues to larger value of $L$. 
Thus it seems to be not ridiculous 
to expect that the asymptotic behavior of the Polyakov loop 
has the oscillating part in the case of $N$ large.

\subsection{Wilson Loop}
   
 We evaluate the Wilson loop amplitude up to the order 1. 
For the operator 
$$
{\cal O}=\frac{1}{4\cdot 3}\sum_{\mu \neq \nu}\frac1N\tr 
(e^{iLX_{\mu}}e^{iLX_{\nu}}e^{-iLX_{\mu}}e^{-iLX_{\nu}}), 
$$ 
we apply the procedure at the beginning of this section. 
The solution of the consistency condition (\ref{self-consistency3}) 
is given by 
\bea
\tilde{\sigma}^2 & = & \sigma^2\left\{1-\frac43\frac{1}{(N+1)!(N-1)!}
e^{-\frac{L^2\sigma^2}{2}}
\left(\frac{L^2\sigma^2}{2}\right)^{2N-2}
+O\left(e^{-L^2\sigma^2}(L^2\sigma^2)^{4N-4}\right)\right\}, \nn \\
\tilde{\tau}^2 & = & \tau^2\left\{1-\frac13\frac{1}{(N+1)!(N-1)!}
e^{-\frac{L^2\sigma^2}{2}}
\left(\frac{L^2\sigma^2}{2}\right)^{2N-2}
+O\left(e^{-L^2\sigma^2}(L^2\sigma^2)^{4N-4}\right)\right\}, 
\eea
with $\sigma^2=\sqrt{\frac{g^2}{N}}$ and $\tau^2=\left(\frac{g^6}{N}\right)^{1/4}$. 
In the limit $L\limit \infty$, $\tilde{\sigma}$ and $\tilde{\tau}$ 
go to the values obtained by the Gaussian 
approximation. 
$\tilde{\lambda}$ is determined as 
\bea
\frac{1}{\tilde{\lambda}} & = & 
\frac12(N^2-1)\left(1-\frac{1}{N^2(N^2+3)}\right) \nn \\ 
 & & +\frac{2}{N!(N-2)!}\left\{-1+\frac{1}{N^2}\left[\frac{2}{3(N+3)}+
2\frac{(N+1)(N+4)}{(N+2)(N+3)}-1\right]
\right\}e^{-\frac{L^2\sigma^2}{2}}\left(\frac{L^2\sigma^2}{2}\right)^{2N-2} \nn \\
 & & +O\left(e^{-\frac{L^2\sigma^2}{2}}(L^2\sigma^2)^{2N-3}\right). 
\eea
After all, we obtain the final result up to the order 1 
\bea
\lefteqn{\bra\hat{W}(L)\ket=\frac{2}{N+1}
\exp\left[\frac{N^2-1}{2N^2(N+3)}\right]} \nn \\
 & \times & 
\left[1-\frac{1}{N!(N-2)!}\frac{2}{N^2}\left\{2\frac{(N+1)(N+4)}{(N+2)(N+3)}-1+
\frac{2}{3(N+3)}\right\}
e^{-\frac{L^2\sigma^2}{2}}\left(\frac{L^2\sigma^2}{2}\right)^{2N-2} \right. \nn \\
 & & \left.\frac{}{}+O\left(e^{-\frac{L^2\sigma^2}{2}}(L^2\sigma^2)^{2N-3}\right)\right]. 
\eea
This behavior has the same tendency as in the bosonic case \cite{OS}. 
Namely, $\bra\hat{W}(L)\ket$ has a nonzero limit as 
$L\limit\infty$, and approaches to this limit from the below. 
The value of the limit $\frac{2}{N+1}\exp\left[\frac{N^2-1}{2N^2(N+3)}\right]$ 
becomes tiny as increasing $N$. 
It is consistent to the numerical result for $L$ larger (up to $L=40$) 
and $N$ smaller ($N=2$, 4, 8) 
in Figure 3(a) in ref. \cite{Krauth}. 
Since our approximation is expected to becomes better for $N$ larger, 
we could say that for the Wilson loop our result nicely 
reproduces the true behavior at least qualitatively.


\section{Discussions}
\setcounter{equation}{0}

  We have considered the reduced four-dimensional supersymmetric Yang-Mills intergral 
with the four 
supercharges by using the Gaussian and mean field approximations. 
We extended the approximation scheme considered in the bosonic Yang-Mills integrals 
so that the supersymmetry is respected. 
As a result, we obtained reasonable results 
for the partition function and the correlators of 
the Polyakov and Wilson loops. 
The result of the partition function shows a nice convergence into the exact value 
as $N$ goes larger. 
Also the Gaussian approximation for the Polyakov and Wilson loops fits 
well the numerical results 
for the region of the loop length $L$ smaller. 
Next we considered the improved version of the mean field approximation 
in order to analyze 
the behavior of the loop amplitudes for $L$ large. 
Although the result of the Polyakov loop does not exhibit good agreement 
to the exact result for the $N=2$ case, it could be expected 
that the situation gets better for 
$N$ large as we saw in the analysis of the partition function. 
The asymptotic behavior of the Wilson loop shows a nice behavior 
that reproduces the numerical result 
at least qualitatively. 

  Comparing with the situation in the bosonic case, we notice these two things. 
For both of bosonic and supersymmetric cases, as $N$ goes larger, 
the result of the partition function tends to converge into the region 
where the approximation 
is reliable. For the supersymmetric case, however it is slower 
than that in the bosonic case. 
Also, in the Gaussian approximation for the Polyakov loop, 
the difference for the region $L\ge 1$ 
between the approximate result and the numerical one  
is larger than in the bosonic case. Thus, we can conclude that 
as $N$ goes smaller and $L$ larger 
the Gaussian approximation gets more worse 
comparing with the result in the bosonic case 
as long as looking at the first few orders. 
Since we have examined the first few terms in the expansion 
around the Gaussian action, 
it is obvious to need investigate the higher order corrections. 
At present, however, computation up to the higher order gets much more 
complicated than that for the first 
few orders. We would need a calculation technique 
making the computation greatly simplified 
or an improvement of the approximation so as 
to give better results up to the first few orders. 

 An interesting direction is an investigation of the case of the supersymmetric 
Yang-Mills integral with eight or sixteen supercharges. 
In this case, there is a technical difficulty 
for obtaining reliable results by numerical simulations 
due to the fermion determinant being not positive 
definite\footnote{There are some attempts to find a 
way out of this problem in refs.\cite{Nishi}.}. 
Since our method seems to be extended also for this case 
by introducing auxiliary fields, 
it would be expected to give concrete results\cite{Sugino2}.

\vspace{3cm}


\begin{large}

{\bf Acknowledgements}

\end{large}

\vspace{7mm}

  The authors would like to thank the authors of ref. \cite{Ambjorn}, 
in particular 
W. Bietenholz and J. Nishimura for sending numerical data 
in the paper and correspondence.

\vspace{3cm}


\end{document}